\documentstyle[12pt]{article}
\def\fun#1#2{\lower3.6pt\vbox{\baselineskip0pt\lineskip.9pt
\ialign{$\mathsurround=0pt#1\hfil##\hfil$\crcr#2\crcr\sim\crcr}}}

\newcommand{\be}{\begin{equation}}
\newcommand{\ee}{\end{equation}}
\newcommand{\bd}{\begin{displaymath}}
\newcommand{\ed}{\end{displaymath}}
\newcommand{\ba}{\begin{array}}
\newcommand{\ea}{\end{array}}
\newcommand{\bt}{\begin{tabular}}
\newcommand{\et}{\end{tabular}}
\newcommand{\bc}{\begin{center}}
\newcommand{\ec}{\end{center}}

\begin{document}

\large

\begin{center}
{\huge\bf Hadroproduction of particle with open charm}
 \footnote{Talk presented at XXXIV Recontres de Moriond "QCD and 
High Energy Hadronic Interactions", Les Arcs, France,  
March 20-27, 1999}\\[0.2cm]

{\large G.H.Arakelyan\\[0.2cm]

 Yerevan Physics Institute, Yerevan 375036, Armenia \\
and\\
Joint Institute for Nuclear Research, Dubna, Russia\\
 E--mails: argev@jerewan1.yerphi.am,~~~~~argev@nusun.jinr.ru
}

\end{center}


\begin{abstract}
In the framework of Quark--Gluon--String Model (QGSM) 
we calculate the inclusive spectra of meson and baryons with 
open charm in hadron--hadron collisions, obtained by SELEX and 
BEATRICE collaborations, taking
into account the decays of corresponding $S$--wave resonances.\\
\end{abstract}
\vspace {1cm}

In the papers \cite{AVZA,AVNAN,AD,AYDS} in the framework of the QGSM 
the inclusive spectra of stable charmed particles 
($\Lambda_c$,$\Xi_c$, $\Omega_c$, $D$, $D_s$) were reanalyzed
taking into account the contributions from decays of corresponding 
$S$--wave resonances, like $1^-$ mesons ($D^*$ and $D^*_s$),
$1/2^+$ ($\Sigma_c$ and $\Xi'_c$) and $3/2^+$ ($\Sigma^*_c$, $\Xi^*_c$
and $\Omega^*_c$) hyperons. $S$--wave charmed resonances decay into 
stable charmed particles with emission of $\pi$--meson or $\gamma$--quantum 
\cite{RPP}. The kinematics of this decays was used according to \cite{ASH}.
In this report our predictions are compared with latest   
experimental data of WA89 collaboration \cite{WA89AS} on 
$\Lambda_c$ charmed hyperon production asymmetry on $\Sigma^-$ beam and 
preliminary measurements of $x_F$ dependence of $\Lambda_c$ charmed 
hyperon cross section and production asymmetry in $\pi^-p$, $pp$ and 
$\Sigma^-p$ collisions of SELEX collaborations \cite{IORY,STENSON,FNAL,KUSH}. 
We also present the comparision of model calculation with the experimetal data 
on vector $D^{*+/-}$ meson production on $\pi^-$- beam measured by 
Beatrice (WA92) collaboration \cite{DSTAR}.
All formulae for inclusive spectra of hadrons together with a full list of 
the quark/diquark distribution functions and corresponding fragmentation 
functions into charmed hadrons used in this work were given in 
\cite{AVZA,AVNAN,AD,AYDS}. We present here only the main features of our 
approach.

The invariant cross section of hadron $h$ production has the form:

\be
\label{1}
x\frac{d\sigma^h}{dx}=x\frac{d\sigma^{h{dir}}}{dx}+
\int\limits^{x^*_+}_{x^*_-}x_R\frac{d\sigma^R}{dx_R}\Phi(x_R)dx_R~.
\ee

Here, $\displaystyle x\frac{d\sigma^{h{dir}}}{dx}$ is the direct production
cross section of hadron $h$, and $\displaystyle x_R\frac{d\sigma^R}{dx_R}$ is
the production cross section of $R$--resonance. Function $\Phi(x_R)$
describes two--body decay of resonance $R$ into hadron $h$. After
integration over transverse momenta of both hadron $h$ and resonance
$R$, the function $\Phi(x_R)$ has the form
\be
\label{2}
\Phi(x_R)=\frac{M_R}{2p^*}\frac{1}{x^2_R}~.
\ee
In eqs. (\ref{1}) and (\ref{2}) $x_R$ is the Feynman variable
of resonance $R$
$$
x^*_+=\frac{M_R\tilde x}{E^*-p^*},\quad x^*_-=\frac{M_R\tilde
x}{E^*+p^*},\quad \tilde x=\sqrt{x^2+x^2_{\bot}},\quad
x_{\bot}=\frac{2\sqrt{<p^2_{\bot}>+m^2}}{\sqrt s},
$$
$m$ is the mass of produced hadron $h$, $M_R$ is the mass of
resonance, $E^*$ and $p^*$ are energy and $3$--momentum of hadron $h$
in the resonance rest frame, $<p^2_{\bot}>$ is the average transverse
momentum squared of hadron $h$.

The hadron $h$ inclusive spectrum in the framework of the QGSM has
the form 
\be
\label{3}
x\frac{d\sigma^h}{dx}=\sum^{\infty}_{n=0}\sigma_n(s)\varphi^h_n(x)~,
\ee
where $\sigma_n(x)$ is the cross section of $n$--pomeron shower
production and function $\varphi^h_n(x)$ determines the contribution of
diagram with $n$ cut pomerons.

The expressions for $\sigma_n(s)$ and values of corresponding parameters 
for $pp$ and $\pi p$ collisions are given elsewhere (see \cite{AVZA} 
for citations).
For $\Sigma^-p$ interaction in the framework of additive quark model we 
calculate the relation between pomeron residues in
$\Sigma^-p$ and $pp$ collisions 
$\displaystyle
\gamma_{\Sigma p}=0.92\gamma_{pp}
$

The functions $\varphi^h_n(x)$ $(n>1)$ for $\pi p$ interaction can be
written in the form \cite{AVZA}
\be
\label{4}
\varphi^h_n(x)=f^h_{\bar q}(x_+,n)f^h_q(x_-,n)+
f^h_q(x_+,n)f^h_{qq}(x_-,n)+2(n-1)f^h_s(x_+,n)f^h_s(x_-,n)
\ee
and for baryon--proton interaction
\be
\label{5}
\varphi^h_n(x)=f^h_{qq}(x_+,n)f^h_q(x_-,n)+f^h_q(x_+,n)f^h_{qq}(x_-,n)+
2(n-1)f^h_{sea}(x_+,n)f^h_{sea}(x_-,n)~,
\ee
where $x_{\pm}=\frac{1}{2}[\sqrt{x^2+x^2_{\bot}}\pm x]$.

The functions $f^h_i(x,n)(i=q,\bar q,qq,q_{sea})$ in (\ref{4}) and
(\ref{5}) describe the contributions of the valence/sea quarks,
antiquarks and diquarks, respectively. They can be expressed as a 
convolution of quark/diquark momentum distribution functions 
$u_i(x,n)$ in the colliding hadrons with the function of
quark/diquark fragmentation into hadron $h$, $G^h_i(x,n)$:
  
\be
\label{6}
f^h_i(x,n)=\int^1_x u_i(x_1,n)G^h _i(x/x_1)dx_1~.
\ee
The projectile (target) contribution depends on the variable $x_+$
$(x_-)$.

For the $\Sigma^-$ beam functions $f^h_q(x,n)$ are
expressed in terms of corresponding $s$-- ($f^h_s(x,n)$) and $d$--quark
($f^h_d(x,n)$) functions in the following form
\be
\label{7}
f^{h(\Sigma^-)}_q(x,n)=\frac13f^{h(\Sigma^-)}_s(x,n)+
\frac23f^{h(\Sigma^-)}_d(x,n)\\[3mm]
\ee

In the framework of the additive quark model diquarks in $S$--wave
baryons may have spin (isospin) $0$ and $1$. Thus the diquark functions
$f^h_{qq}(x)$ are expressed in terms of scalar $(0)$ and vector $(1)$
diquark functions with the weights determined by $SU(6)$ symmetric
functions
\be
\label{8}
f^{h(p)}_{qq}=\frac13f^{h(p)}_{uu}(x,n)+\frac12f^{h(p)}_{(ud)_0}(x,n)+
\frac16f^{h(p)}_{(ud)_1}(x,n)\\[3mm]
\ee
\be
f^{h(\Sigma^-)}_{qq}=\frac13f^{h(\Sigma^-)}_{dd}(x,n)+
\frac12f^{h(\Sigma^-)}_{(ds)_0}(x,n)+
\frac16f^{h(\Sigma^-)}_{(ds)_1}(x,n)\\[3mm].
\ee

In what follows, we will assume that the distribution functions of scalar
and vector diquarks $u_{qq}(x,n)$ are the same. Certainly, different
diquarks  fragment into baryons in different ways: for
instance, the direct production of $\Lambda_c$ in $pp$ collision is
determined by scalar (and isoscalar) diquark function $f_{(ud)_0}$
 and direct production of $\Sigma_c$ and $\Sigma^*_c$
hyperons is determined by the vector diquark function $f_{(ud)_1}$.
 

Further, we will assume that the
fragmentation functions of quarks and diquarks do not depend on
spin of the picked up quark (or diquark). In particular this means that 
functions for fragmentation into $\Sigma_c$ and $\Sigma^*_c$, 
$D$ and $D^*$ mesons are equal.

Let now turn to the consideration of the experimental data on $L_c$ 
production measured by SELEX collaboration on different hadron beams. 

The unnormalized inclusive spectra of $\Lambda_c$--baryon in $\pi^-p$, 
$pp$ and $\Sigma^-p$ collisions at $P_L=600GeV/c$  were compared 
with model calculations on Fig.1a-c respectively. 
We present here two set of the data: full circles -- the first publication 
of SELEX data \cite{IORY} and full squares -- the new one, presented by 
K.Stenson on this conference \cite{STENSON} and also published in \cite{FNAL}. 
The theoretical curves 
are calculated taking into account both direct $\Lambda_c$ production  
and the one which is due to the contribution of $\Lambda_c$ produced in 
decays $\Sigma_c\to\Lambda_c\pi$ and $\Sigma^*_c\to\Lambda_c\pi$ with 
free parameters, given in \cite{AVZA} exept the parameter, standing for 
$\bar L_c$ contribution. The value $a4=0.1$ was taken to have a satisfactory 
description of $A(L_C,\bar L_c)$ production asymmetry(see below). As one 
can see the model reproduce the shape of the inclusive spectra rather 
well. Concerning the data on $\Sigma^-p$ collisions (Fig.1c) the 
disagreement at small $x_F$ in the first set of the data \cite{IORY} may be 
due to experimental uncertainty and as we can see from Fig.1c the new set of the 
data \cite {STENSON,FNAL} do not have such 
strong decrease at small $x_F$. Note, that as in \cite{AVZA} the theoretical 
curves were calculated without taking into account the charmed sea contribution.


In Fig.2 we plotted the experimental points of the $L_c$ and 
$\bar L_c$ production asymmetry measured by two collaborations WA89 at 
$P_L=340GeV/c$ \cite{WA89AS}(full circles) and SELEX \cite{KUSH} at 
$P_L=600GeV/c$ (stars) in $\Sigma^-p$ collisions together with QGSM 
prediction.
Both experimental data of WA89 \cite{WA89AS} and SELEX \cite{KUSH} shows 
rather large production asymmetry even at $x_F>0.2$.  Two curves were 
calculated for momenta 
$P_L=340GeV/c$ (full line) and $P_L=600GeV/c$ (dashed line). The difference 
at small $x_F$ has a kinematical reason. At large value of $x_F$ the 
theoretical curves are energy independent. As we can see the modified QGSM 
calculations reproduce the experimental behavior both inclusive cross sections 
and production asymmetry on different hadron beams.

Let now turn to the description of the charmed vector meson production. 

It is important for the model under consideration to compare our 
predictions with spectra of resonances. The data on inclusive 
differencial cross section $1/\sigma d\sigma/dx$ of the reactions 
$\pi^-p\to D^{*+}/D^{*-}X$ and  
production asymmetry $A(D^{*-}/D^{*+})$ at $350\;GeV/c$ \cite{DSTAR} 
together with theoretical predictions are shown in Figs.3a,b and Fig.4 
respectively. 


The theoretical calculations were perform  with taking into 
account the charmed sea contribution described in \cite{AD}. All 
formulae for the parametrisations of the $c$-sea in the initial 
pion and the $c$ quark fragmentation functions into $D^*$ mesons 
were given in \cite{AD}.

The full line (1) corresponds to 
calculations without taking into account charmed sea contribution, dashed (2)
and dashed-dotted (3) lines stands for different parametrisations of leading 
fragmentation functions into $D^{*-}$ meson as described in \cite{AD} 
( formulae (14) and (15) correspondingly).


It seems that the agreement for the shape of the sum of 
spectra of $D^{*+}$ and $D^{*-}$ mesons is quite satisfactory.

The production asymmetry of the leading
 ($D^{*-}, \bar D^{*0}$) and nonleading ($D^{*+}, \\
 D^{*0}$) vector
resonances with open charm in  $\pi^- p$-- collisions, integrated
over the $x_F>0$ region was measured in \cite{ALPRD49}. The experimental 
value of asymmetry was obtained equal $A(L,NL)=0.09 \pm 0.06$.

The calculations of the model give the following values for all
parametrizations used in work \cite{AD}:
Variants 1,2 and 3 correspods to the different parametrisation 
of the fragmentation functions as described above.
The behaviour of production asymmetry for $D^{*-/+}$ vector mesons 
presented on Fig.4 look like that for the pseudoscalar $D$ mesons. 
It is important to note that the measurements of the integrated 
cross section of $D^{*+}$ and $D^{*-}$ mesons are about twice lower 
than data \cite{ABPL169} measured on hydrogen target at the same 
momenta.

{\it Acknowledgements}. The author is grateful to A.B.Kaidalov and 
K.G.Bo-reskov for useful discussions. I express my gratitude to 
C.Lazzaroni (WA92 BEATRICE collaboration) for giving the preliminary 
data of collaboration and useful discussions. 
I also thank M.Iori (SELEX) for preliminary data on $L_c$ cross section 
reported  on Genova conference and also to G.Davidenko, A.Nilov and 
F.Garcia for a useful discussions of the SELEX data on $L_c$ production.

This work was supported in part by  NATO grant OUTR.LG 971390
and RFBR grant 98-02-17463.

\newpage

\section*{Figure captions}

\noindent
\bt{p{1.5cm}p{14.5cm}}

Fig.1 & Comparison of QGSM calculations with preliminary experimental 
data on $\Lambda_c$ spectra measured by SELEX collaboration \cite{IORY,
STENSON,FNAL}
for :a) $\pi^-p$ , b) $pp$ and c) $\Sigma^-p$ collision at $P_L=600GeV/c$.\\

Fig.2 & The $x_F$--dependence  of $L_c, \bar L_c$--production asymmetry 
in $\Sigma^-p$ collision. The experimental data from WA89 \cite{WA89AS} 
(full circles) at $P_L=340GeV/c$ and SELEX \cite{KUSH}(open circles) 
at $P_L=600GeV/c$.The theoretical curves were calculated for 350GeV/c 
(full line) and 600GeV/c(dashed line).\\

Fig.3 & The $x_F$--dependence of $D^*$--meson cross section  
$1/\sigma d\sigma/dx$ in $\pi^-p$ interaction at $350\;GeV/c$ 
\cite{DSTAR}\\

Fig.4 & Comparison of the model calculations with experimental data on
leading ($D^{*-}$) and nonleading ($D^{*+}$) charmed vector mesons 
production asimmetry in $\pi^-p$ interaction at $P_L=350\;GeV/c$
\cite{DSTAR}.\\
\et

\newpage


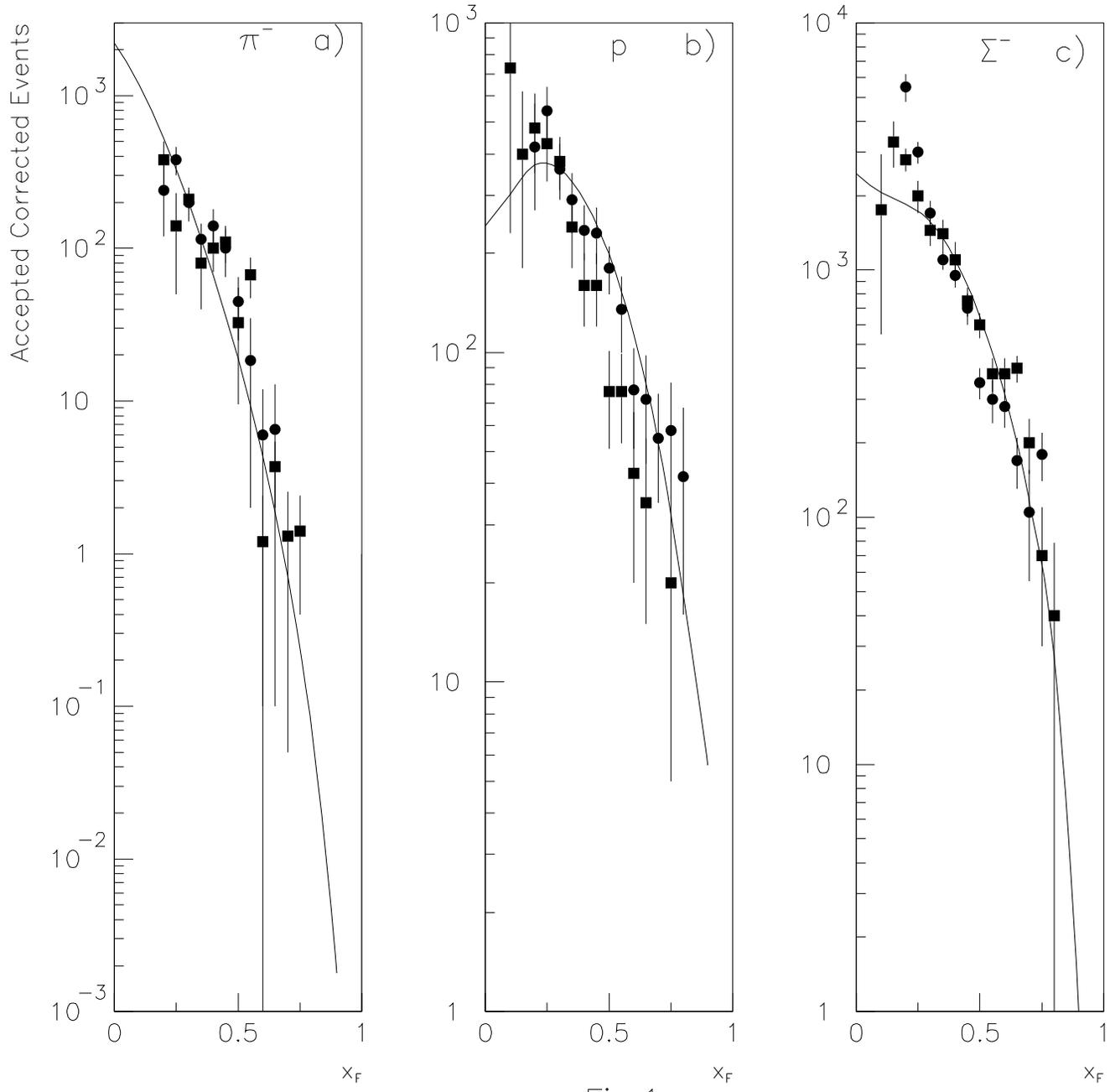
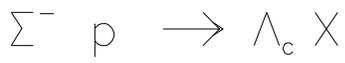
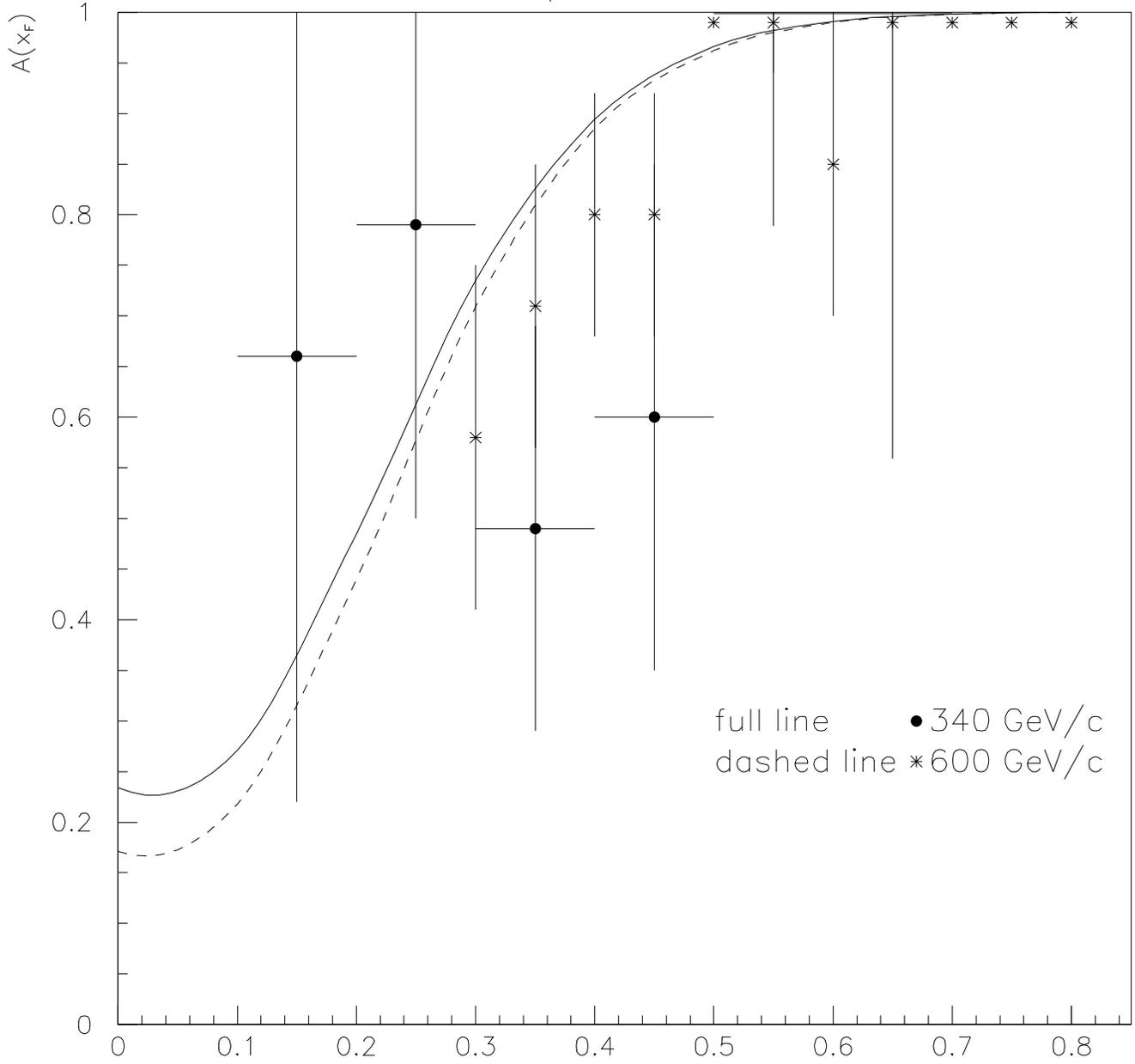
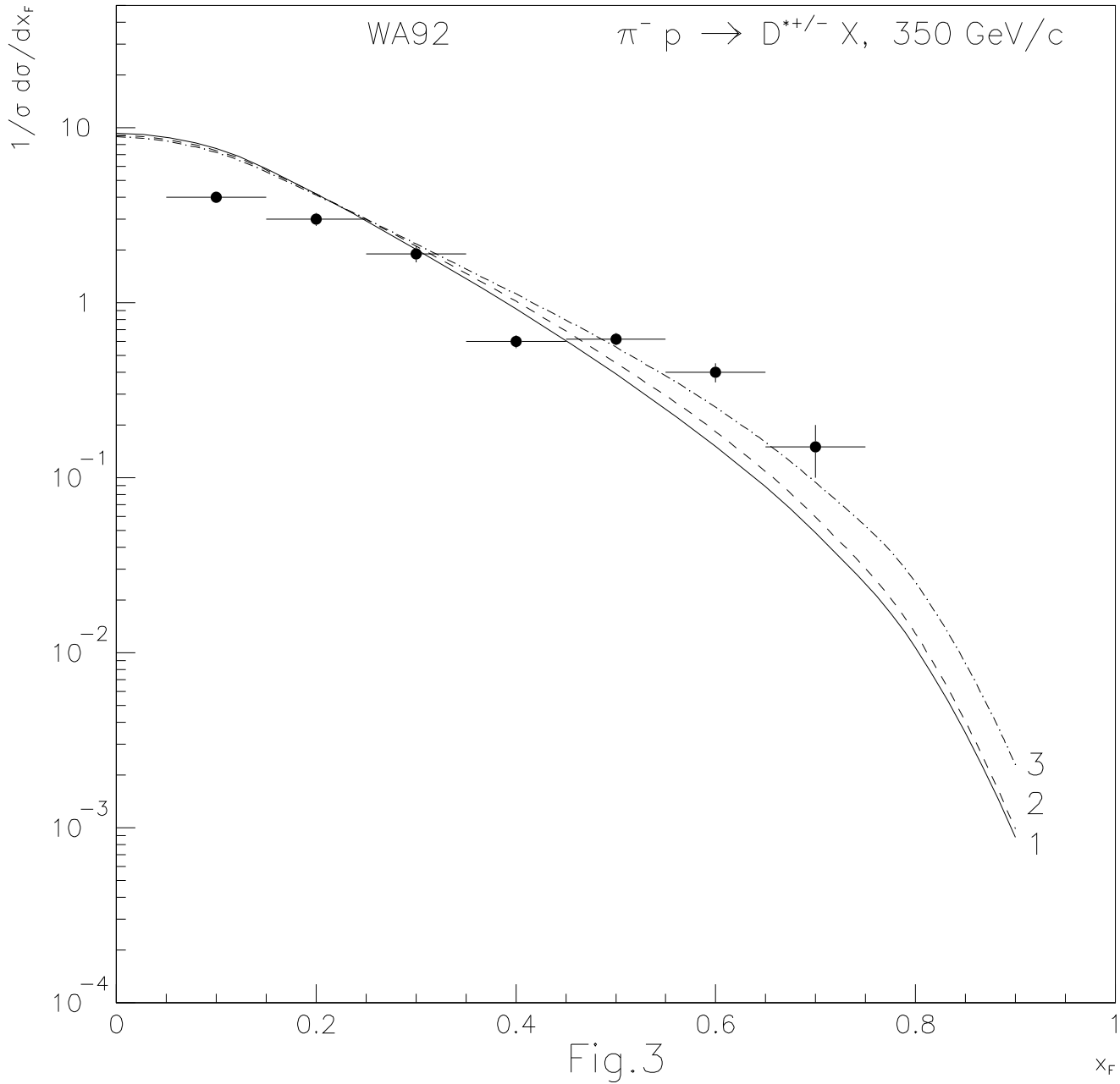
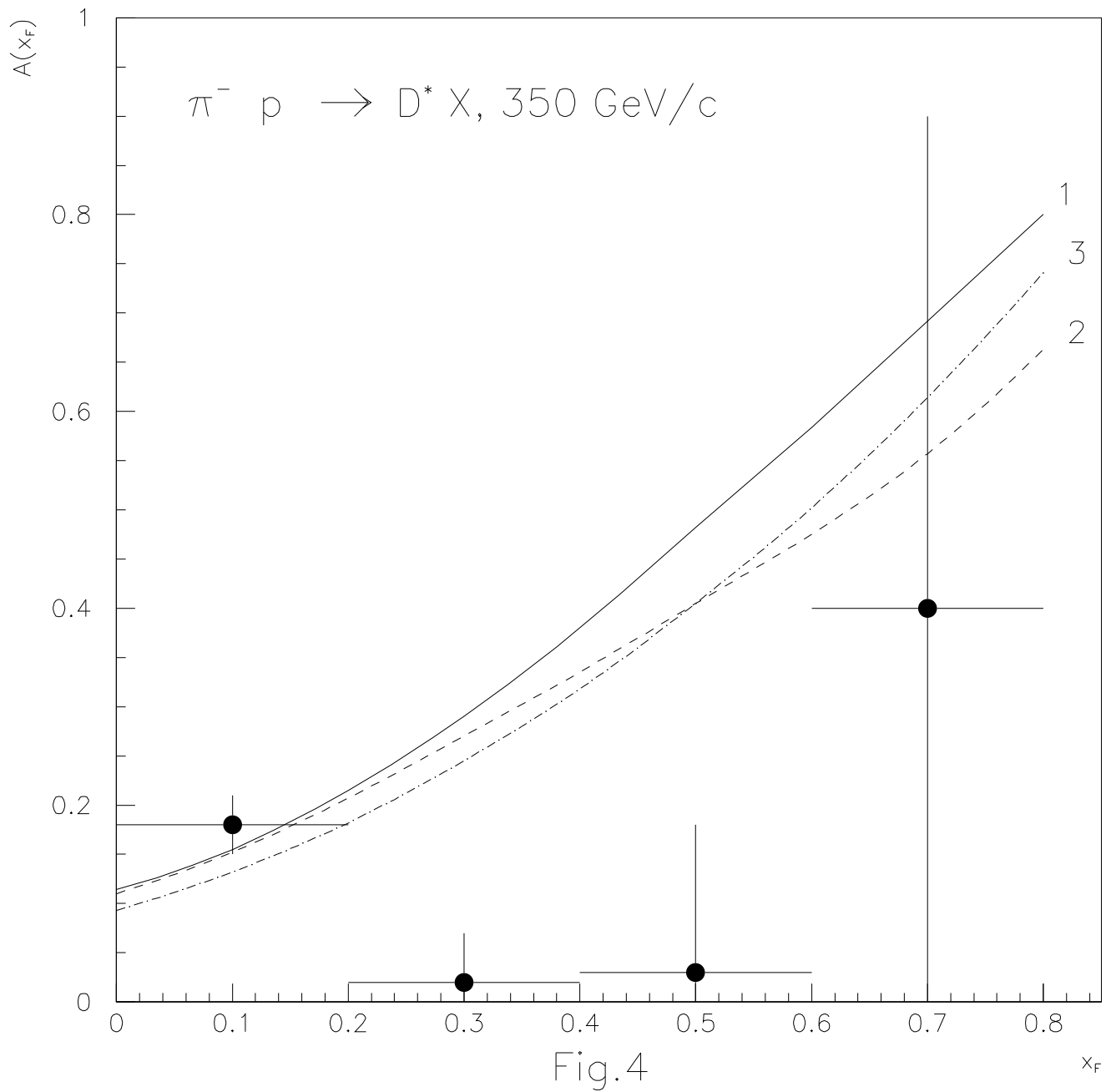
\end{document}